\documentclass[aps,pra,epsfigure,twocolumn]{revtex4}
\usepackage{amsmath}
\usepackage{amstext}
\usepackage{latexsym}
\usepackage{graphicx}
\usepackage{amsfonts}
\usepackage{epsfig}
\usepackage{color}
\usepackage[abs]{overpic}

\newcommand{\Tr}{\mathrm{Tr}}

\begin{document}
\title{Noise powered amplification in CV quantum key distribution}
\author{Petr Marek and Radim Filip}
\affiliation{Department of Optics, Palack\' y University,\\
17. listopadu 1192/12,  771~46 Olomouc, \\ Czech Republic}
\begin{abstract}
Amplification plays a key role in classical communication protocols, where it compensates the unavoidable loss of the signal. However, when we enter the quantum domain this approach starts being problematic as the standard kinds of amplifiers are usually accompanied by excess noise which detrimentally affects the quantum features of used states. Recently, several kinds of ``noiseless'' amplifiers that do not suffer from this feature have been proposed. Among these amplifiers, one that stands out is the noise powered amplifier, which acts incoherently, amplifying the signal by adding the ``right'' kind of noise.  Here we show that despite the incoherence, which makes the amplifier unsuitable for protocols such as entanglement distillation, the added noise is not a big problem in quantum key distribution tasks and can be in some situations beneficial.
\end{abstract}
\maketitle

\section{Introduction}
Quantum physics poses hard fundamental limits on our ability to manipulate with information encoded into physical systems. One of these is the inability of extracting all of this `quantum information' just from a measurement on a single copy of the quantum system, which is strongly related to the well known no-cloning theorem \cite{nocloning}. It is also the driving principle behind quantum key distribution (QKD) - the most developed application of quantum information theory yet. Continuous variable (CV) QKD \cite{CVQKD1} has recently appeared as an alternative to the initially developed protocols with single photons. It utilizes coherent and squeezed states of light, relying on homodyne instead of single photon detectors. In the absence of noise it principally allows for secure communication over arbitrarily lossy channel \cite{CVQKD2}. In reality, though, certain amount of noise is unavoidable and loss is therefore the chief factor in limiting the communication distance.

In classical communication protocols it is customary to compensate the loss with help of amplifiers - an approach which not possible in the quantum domain, as the existing phase insensitive amplifiers add enough noise to the system to render whole procedure counterproductive \cite{Caves82}. In fact, the ideal noiseless amplifier, which would be able to fully compensate the loss at least at some level by allowing coherent states to be transformed as $|\alpha\rangle \rightarrow |g\alpha\rangle$, is represented by an unbounded trace increasing operator and therefore does not exist as a physical operation.

However, it is possible to devise an approximation of the amplification device. There have been several proposals with varying levels of technological sophistication, ranging from proposals based on quantum scissors \cite{scissors}, over additions and subtractions of individual photons \cite{marekamp, HFteor, HFexp}, up to deliberate injection of noise followed by photon subtractions \cite{marekamp, NPexp}. The last mentioned kind of amplifier is different from the others - it utilizes a noise addition as part of the process. Noise addition, which is exactly the opposite of that what we are trying achieve.  This is not without repercussions, as this type of amplifier is, unlike the others, incoherent. It reduces purity of the state and, when applied to entangled state, reduces entanglement as well. Interestingly enough, quantum communication is not only about entanglement and in this paper we show, there are regimes in which this apparently less-than-ideal amplifier outperforms its more effective counterparts.

Recently there have been several studies related to application of amplifiers in communication tasks. In \cite{ampralph} it was shown that the amplifier based on quantum scissors can be used for compensation of losses and distillation of entanglement. A related issue was discussed in \cite{virtualamp1,virtualamp2}, where it was shown that Gaussian data processing virtually simulating the effect of the ideal amplifier can be used to enhance the rate of secure key distribution in CV QKD.

In this paper we look closely at two kinds of noiseless amplifiers and analyze how beneficial they can be when used in communication protocols. In Sec.~\ref{Sec2} we introduce the formal description of the communication protocol and of the amplifiers. In Sec~\ref{Sec3} we analyze the amplifiers for use in communication over ideal channels. In sec.~\ref{Sec4} we expand the treatment by considering realistic channels with loss and noise. The results are finally summarized in sec.~\ref{Sec5}.

\section{Communication protocol and amplification}\label{Sec2}
\begin{figure}
\centerline{\includegraphics[width = 8cm]{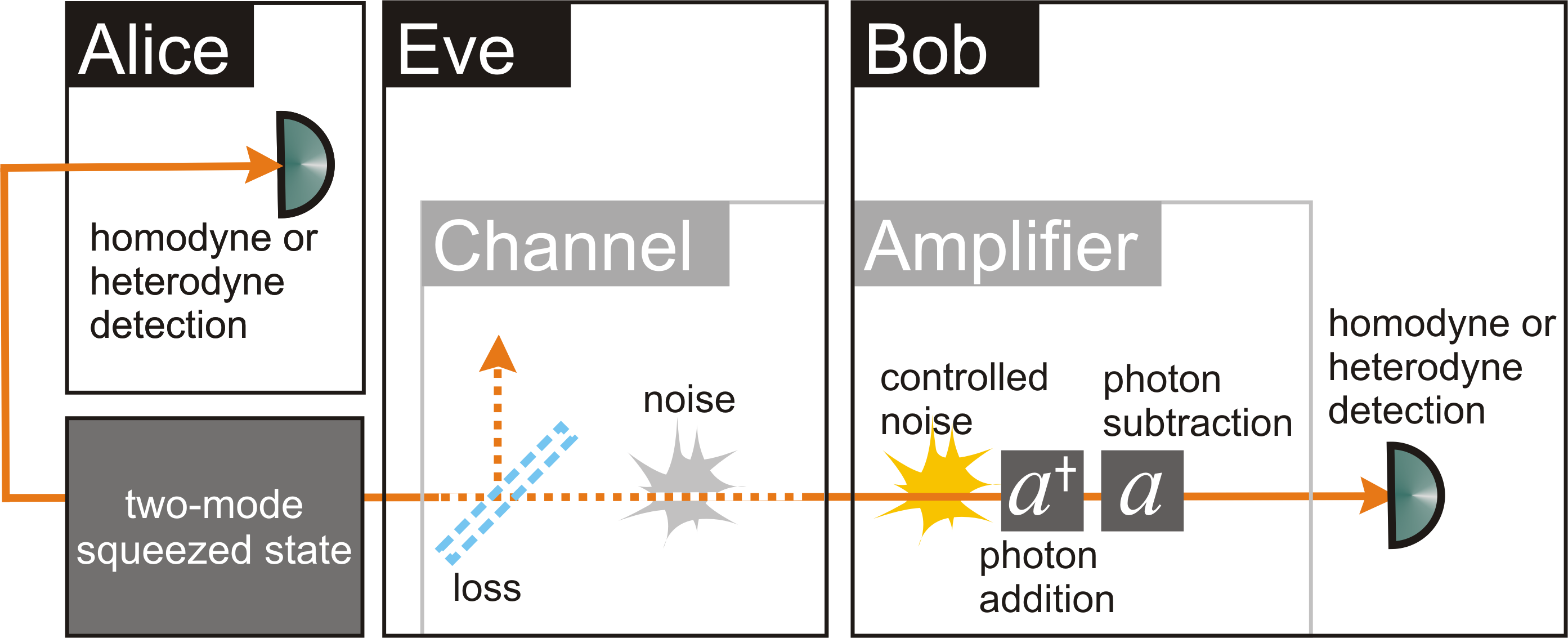}}
\caption{(Color online) Schematic setup of the communication protocol. }
\label{figure_setup}
\end{figure}

The secure key distribution between Alice and Bob, which is schematically depicted in Fig.~\ref{figure_setup}, can be always represented by an entangled two-mode quantum state, which is shared and locally measured by the two communicating parties. Entanglement is a necessary part of the state, without it the shared correlations are inherently classical and no secure key can be established. In a continuous variables CV quantum key distribution, the shared quantum state is a two mode squeezed state and the measurements performed can be either of the homodyne, or of the heterodyne variety. The measurement at Alice's side determines the nature of the communication: homodyne or heterodyne measurement represents communication with squeezed or coherent states, respectively. Measurement at Bob's side bears no such significance.

There are several factors deciding the amount of information which can be transmitted in this way: the shared quantum state and the measurements performed on its two parts. For two sets of measurements, $A$ and $B$, with possible measurement outcomes $\{a\}$ and $\{b\}$, respectively, the amount of mutual information shared by Alice and Bob after a single round of measurements on quantum state $\rho$ is given by:
\begin{equation}\label{IAB}
    I(A,B) = \int p(a,b) \log_2\frac{p(a,b)}{p(a)p(b)} da db,
\end{equation}
where $p(a,b) = \mathrm{Tr} [\rho \Pi_a \otimes \Pi_b ]$ is a joint probability distribution of Alice measuring value $a$ and Bob measuring value $b$, where these events are represented by the projectors $\Pi_a$ and $\Pi_b$. $p(a)$ and $p(b)$ are then marginal probability distributions of the individual measurements. For our analysis we consider two possible classes of measurements, which have been previously applied to CV quantum communication. Homodyne measurement returns single real values $x$, each of which corresponds to one particular projector on a specific $x$ quadrature eigenstate $|x\rangle\langle x|$. Heterodyne measurement consists of two homodyne measurements measuring in two conjugated bases, as such it results in a pair of values $x$ and $p$, with corresponding projector $\Pi_{x,p} = |\alpha\rangle\langle \alpha|$, where $|\alpha\rangle$ is a coherent state with amplitude $\alpha = (x+ ip)/\sqrt{2}$.
In principle, any non-factorable quantum state shared between Alice and Bob can be used for communication. Practical CV communication protocols employ exclusively Gaussian states, fully described by their variance matrix, for which the formulas can be simplified significantly \cite{CVQKD2}.

In the ideal scenario, the shared quantum state $\rho$ is pure and the amount of mutual information is given by its entanglement. This ceases to be the case in the presence of channel loss and noise. However, their influence only transforms the density matrix of the shared state and the methodology for obtaining the amount of information transferrable remains unchanged. The same goes for any kind of probabilistic operation applied to the quantum state. In the following we shall be mostly concerned with various kinds of noiseless amplification, which aims at increasing the mutual information at the cost of reduced probability of success. Operations of these kind can be always represented by trace decreasing maps $\mathcal{A}$ and the amplified state is then
\begin{equation}\label{Amp1}
    \rho_{amp} = \frac{\mathcal{A}(\rho)}{\mathrm{Tr}[\mathcal{A}(\rho)]}.
\end{equation}
Ideal noiseless amplification is represented by operator $G = g^n$, where $n$ is the photon number operator, and it transforms an arbitrary coherent state into $G|\alpha\rangle = |g \alpha \rangle$. The operation is easy to work with and it has the added benefit of being Gaussian - it preserves the Gaussian nature of Gaussian states, which are the staple of contemporary communication protocols. However, since the operator is unbounded for $g>1$, it does not represent a physical operation and in its exact form it can not be experimentally implemented. When one is interested in amplification of weak states, though, the amplification can be approximatively implemented in several ways \cite{}. It should be stressed that all these realistic amplification procedures are not Gaussian, one can therefore no longer rely just on the description by variance matrix and needs to employ the density matrix to obtain the mutual information (\ref{IAB}).

In the following, we shall focus on amplifiers employing the photon-addition and photon-subtraction operations \cite{marekamp, HFteor}, performance of which has been experimentally verified for coherent states with reasonably large amplitudes \cite{NPexp, HFexp}. The two amplifiers are the `High-Fidelity amplifier' (HFA) and the 'Noise-powered amplifier' (NPA). The first one can be implemented by adding-and-then-subtracting a number of photons, while the second one replaces the photon addition step by deliberate addition of thermal noise. As a consequence, HFA preserves purity of the amplified quantum state and can be, in principle, employed for entanglement distillation, while NPA can not. On the other hand, NPA is much less experimentally demanding significantly higher numbers of subtracted photons can be achieved.

We can devise a general form fusing HFA and NPA together. Such the amplifier is obtained by the sequence of noise addition, photon addition, and photon subtraction, and it transforms an arbitrary quantum state $\rho$ to
\begin{equation}\label{Amp2}
   \frac{1}{\mathcal{N}} a^N a^{\dag M} \left[\int \frac{e^{-|\alpha|^2/\Delta}}{\pi\Delta} D(\alpha)\rho D^{\dag}(\alpha) d^2\alpha\right] a^M a^{\dag N},
\end{equation}
where $\mathcal{N}$ is a normalization factor assuring unit trace of the resulting density matrix. The amplifier is quantified by three parameters - the amount of added noise $\Delta$ and the numbers of photons added, $M$, and subtracted $N$. HFA and NPA are now  obtained by setting $\Delta = 0$ and $M = 0$, respectively.

\begin{figure}
\centerline{\includegraphics[width = 5.5cm]{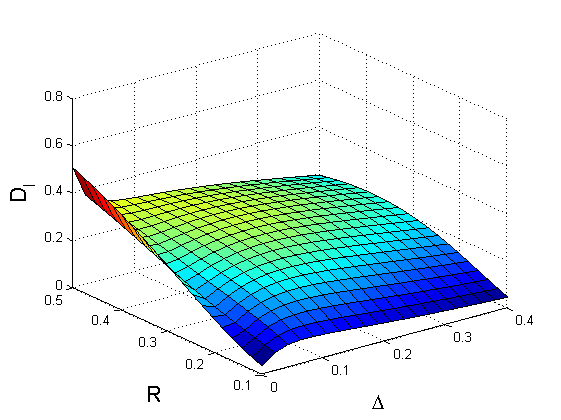}}
\vspace{-4cm}
\centerline{{\bf (a)}}
\vspace{3.5cm}
\centerline{\includegraphics[width = 5.5cm]{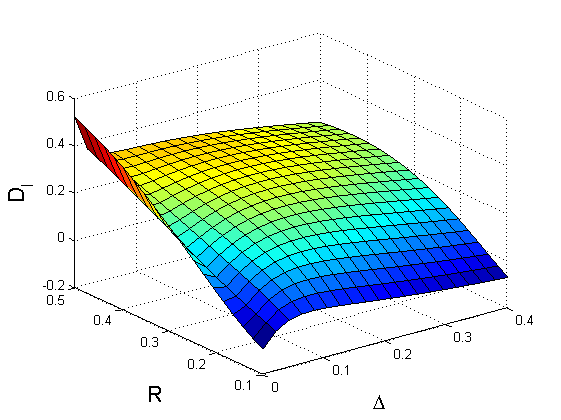}}
\vspace{-4cm}
\centerline{{\bf (b)}}
\vspace{3.5cm}
\centerline{\includegraphics[width = 5.5cm]{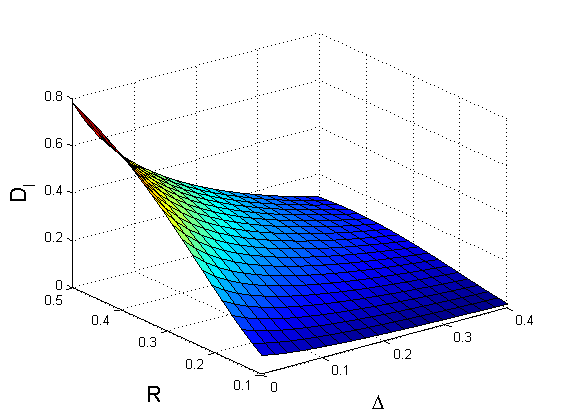}}
\vspace{-4cm}
\centerline{{\bf (c)}}
\vspace{3.5cm}
\caption{(Color online) Difference of mutual information for two-photon subtraction (a), three photon subtractions (b), and addition and subtraction of a photon (c), in relation to the squeezing of the initial state $R$ and the amount of noise added prior to the operations $\Delta$. Both Alice and Bob perform homodyne detection.  }
\label{figure11}
\end{figure}
\section{Ideal scenario}\label{Sec3}
Let us first analyze the ideal scenario, in which the channel is perfect - there is no loss or noise present in it. This corresponds to situation in Fig.~\ref{figure_setup} only with Eve completely removed from the picture. In this case the mutual information is solely given by the entanglement of the initial two-mode squeezed state, which is a Gaussian state fully described by its variance matrix
\begin{equation}\label{varmatrix}
V = \frac{1}{2}\left(\begin{array}{cccc}
      \cosh 2R & 0 & \sinh 2R & 0 \\
      0 & \cosh 2R & 0 & -\sinh 2R \\
      \sinh 2R & 0 & \cosh 2R & 0 \\
      0 & -\sinh 2R & 0 & \cosh 2R
    \end{array}\right),
\end{equation}
where $R$ is the squeezing parameter. In Fock representation, the density matrix of the state can be expressed as
\begin{equation}\label{}
    \rho = \frac{1}{\cosh 2R} \sum (\tanh 2R)^{n+m} |n,n\rangle\langle m,m|.
\end{equation}
From here we can directly obtain the amplified form of state by using (\ref{Amp2}), and then use this new amplified state to obtain new mutual information shared by the communicating parties. We can then plot the difference of mutual information before and after the amplification, $D_I = I(A,B)-I_{0}(A,B)$ in dependence on the parameters of the communication. One of these parameters is the squeezing parameter $R$ representing the communication, the other is the amount of noise added in the amplification step $\Delta$.
\begin{figure}
\centerline{\includegraphics[width = 5.5cm]{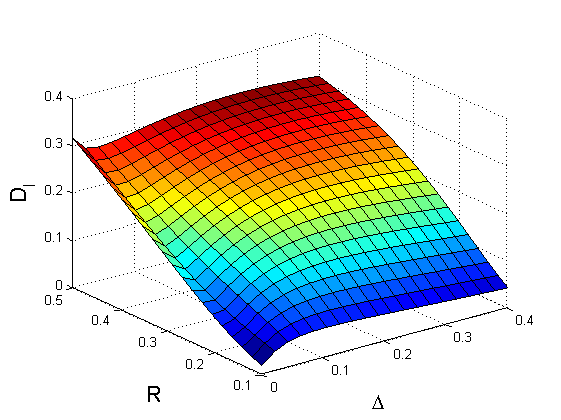}}
\vspace{-4cm}
\centerline{{\bf (a)}}
\vspace{3.5cm}
\centerline{\includegraphics[width = 5.5cm]{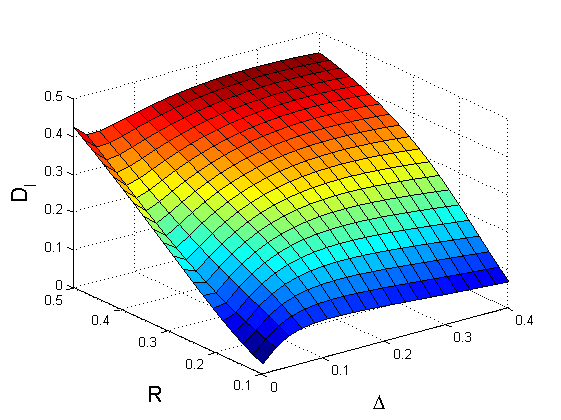}}
\vspace{-4cm}
\centerline{{\bf (b)}}
\vspace{3.5cm}
\centerline{\includegraphics[width = 5.5cm]{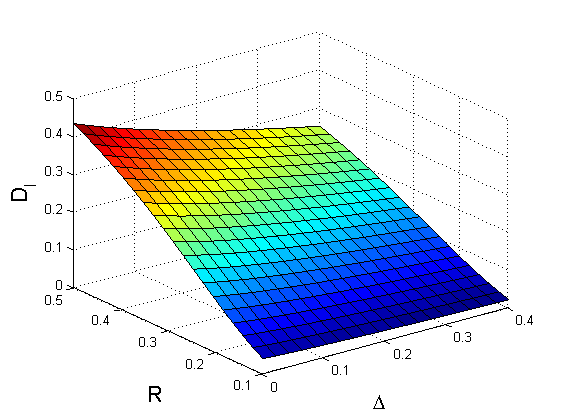}}
\vspace{-4cm}
\centerline{{\bf (c)}}
\vspace{3.5cm}
\caption{(Color online) Difference of mutual information for two-photon subtraction (a), three photon subtractions (b), and addition and subtraction of a photon (c), in relation to the squeezing of the initial state $R$ and the amount of noise added prior to the operations $\Delta$. Alice and Bob perform homodyne and heterodyne detections, respectively.  }
\label{figure12}
\end{figure}

The three specific amplification configurations we are going to compare are: addition of noise and subtraction of two photons, addition of noise and subtraction of three photons, and addition of noise followed by addition and then subtraction of a photon. The first two are essentially the same device, differing only by the number of subtractions, which is responsible for strength of the operation. The third one is a hybrid of the pure HFA and NPA. Values for the ideal HFA are easily obtained by looking at the edge of the graph, but the full picture will be useful in determining whether there are any benefits to adding noise.
\begin{figure}
\centerline{\includegraphics[width = 5.5cm]{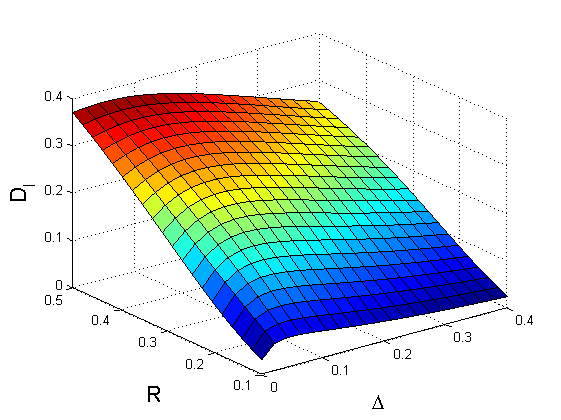}}
\vspace{-4cm}
\centerline{{\bf (a)}}
\vspace{3.5cm}
\centerline{\includegraphics[width = 5.5cm]{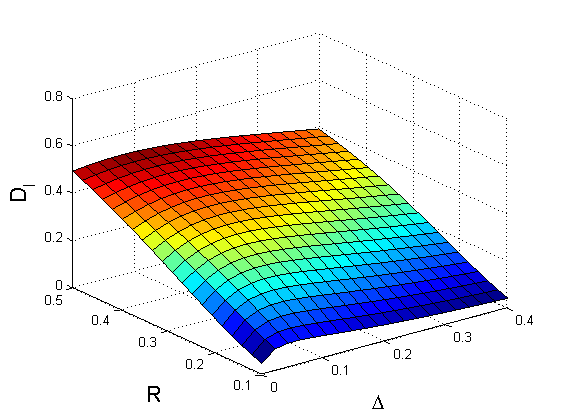}}
\vspace{-4cm}
\centerline{{\bf (b)}}
\vspace{3.5cm}
\centerline{\includegraphics[width = 5.5cm]{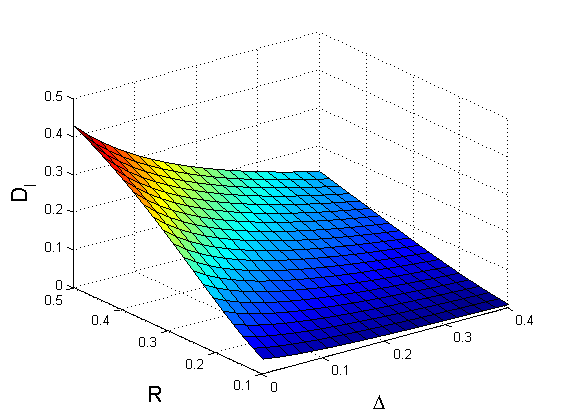}}
\vspace{-4cm}
\centerline{{\bf (c)}}
\vspace{3.5cm}
\caption{(Color online) Difference of mutual information for two-photon subtraction (a), three photon subtractions (b), and addition and subtraction of a photon (c), in relation to the squeezing of the initial state $R$ and the amount of noise added prior to the operations $\Delta$. Alice and Bob perform heterodyne and homodyne detections, respectively. }
\label{figure21}
\end{figure}

In Fig.~\ref{figure11} we see the difference in mutual information when both Alice and Bob perform homodyne detections. This corresponds to direct communication, in which Alice encodes information into a displaced squeezed state and Bob measures the displacement by homodyne detection. We can see that all three nonclassical operations we have considered, two photon subtraction in Fig.~\ref{figure11}a, three photon subtraction in Fig.~\ref{figure11}b, and photon addition and subtraction in Fig.~\ref{figure11}c, increase the mutual information when no noise is present, with the last one of the three providing the largest benefit. In this scenario, the effect of added noise is almost completely detrimental.  The exception are the photon subtraction schemes, where a larger amount of noise is usually better than a small one (even though no noise is the best, unless the shared entanglement is very small).
\begin{figure}
\centerline{\includegraphics[width = 5.5cm]{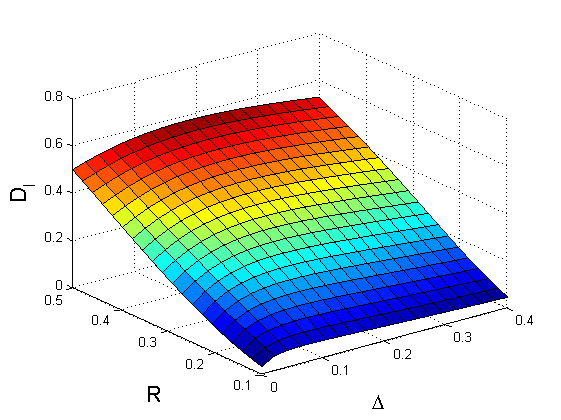}}
\vspace{-4cm}
\centerline{{\bf (a)}}
\vspace{3.5cm}
\centerline{\includegraphics[width = 5.5cm]{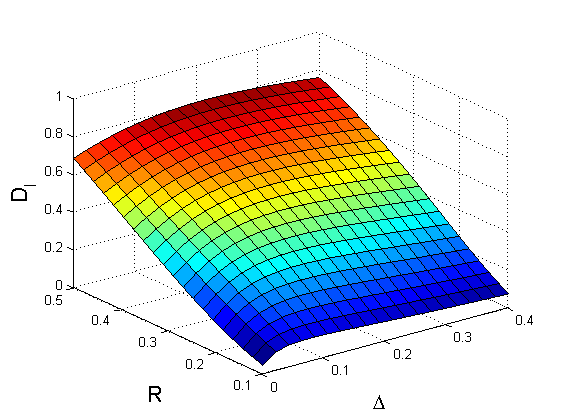}}
\vspace{-4cm}
\centerline{{\bf (b)}}
\vspace{3.5cm}
\centerline{\includegraphics[width = 5.5cm]{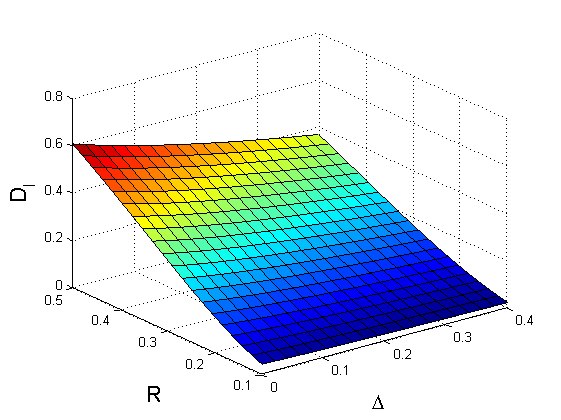}}
\vspace{-4cm}
\centerline{{\bf (c)}}
\vspace{3.5cm}
\caption{(Color online) Difference of mutual information for two-photon subtraction (a), three photon subtractions (b), and addition and subtraction of a photon (c), in relation to the squeezing of the initial state $R$ and the amount of noise added prior to the operations $\Delta$. Both Alice and Bob perform heterodyne detections. }
\label{figure22}
\end{figure}

The situation starts being different when Bob uses heterodyne measurement instead of the homodyne one, while Alice keeps using homodyne detection (encoding into squeezed states), which is the situation represented by Fig.~\ref{figure12}. This is slightly similar to a situation in in CV QKD, where the seemingly suboptimal measurement on the Bob's side allows for increasing secure communication distance by reducing information gained by eavesdropper Eve \cite{heteroQKD}. In this scenario all the employed non-Gaussian operations lead to improvement of mutual information. And again, when only photon subtractions are used, addition of noise can improve the information gain even further. And this can be said for all regimes, in which coherent states are employed. Next to the squeezed-heterodyne protocol, which was just mentioned, there is the encoding into coherent states, which are measured either by homodyne, or by heterodyne detection. This scenarios are represented by Fig.~\ref{figure21} and Fig.~\ref{figure22}.

In summary we can say that, when communicating solely with squeezed states and homodyne detection, the non-gaussian operations lead to improvement, which is generally not helped by inclusion of additional noise. However, as soon as coherent states start being employed, either at the state preparation or at the measurement stage, the situation changes. While photon addition and subtraction still works best without any noise present, photon subtraction schemes are further enhanced by it. It can be even said that the noise compensates the advantage photon addition has over photon subtraction. On the level of mutual information, addition and subtraction of photon is observably better than subtraction of two photons and roughly on the level of three photon subtractions. With added noise, two subtractions can match the addition and subtraction, and three subtractions overcome it significantly.

\begin{table*}
\label{holevo_comparison}
\includegraphics[width = 0.6\linewidth,natwidth=1747bp,natheight=2586bp]{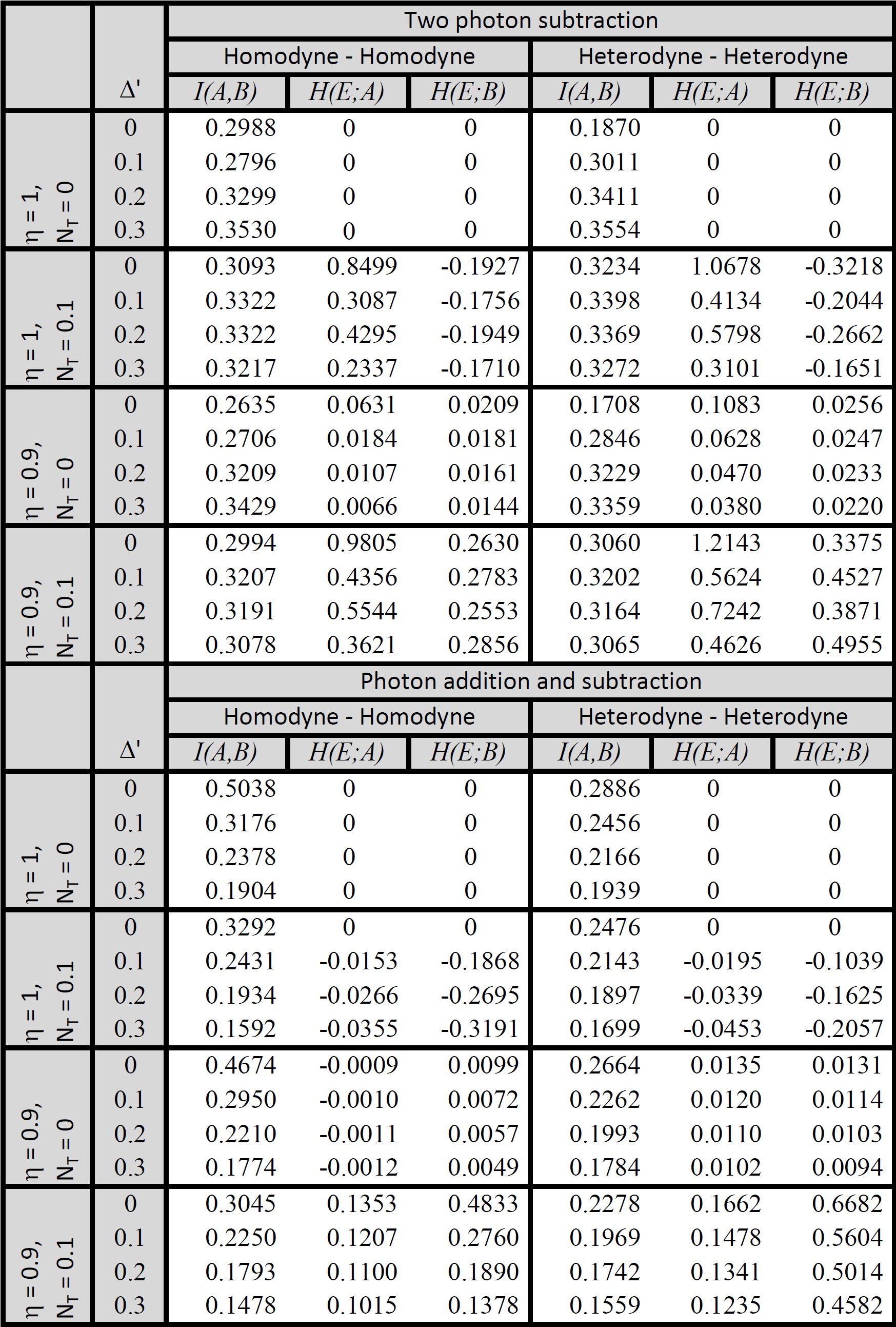}
\caption{Comparison of differences in mutual information between Alice and Bob, $I(A,B)$, and differences Holevo quantities for the eavesdropper Eve for both direct and reverse reconciliation, $H(E;A)$ and $H(E,B)$, respectively. The values are compared across two kinds of amplification operations with several levels of added noise, two communication protocols characterized by measurements performed by Alice and Bob, and four different channels represented by their loss $\eta$ and added noise $N_T$. See the main text for more details. }
\end{table*}

\section{Realistic channel}\label{Sec4}
So far we have only considered the ideal case, in which the communication channels suffers from no losses or noise. This is an artificial scenario far removed from practical conditions. It's results give us some insight into the benefits of the amplification, but it is not enough. The biggest concern related to the loss and noise is not about how information between Alice and Bob gets lost, but how it leaks out towards the third party - the nefarious eavesdropper Eve. Increasing mutual information of Alice and Bob is of no use, if the information Gained by Eve is raised by the same, or higher, amount. We shall not restrict Eve to using a specific type of eavesdropping interaction or measurements. In our treatment by effective entanglement, we can represent the communication by a three-mode entangled state $\rho_{ABE}$ such that $\rho_{AB} = \Tr_E[\rho_{ABE}]$ is equivalent to a density matrix of a two-mode squeezed state for which one of the modes is affected by loss $\eta$ and noise $N_T$. As a consequence, $\rho_{AB}$ represents a Gaussian state with variance matrix
\begin{equation}\label{}
    V_{AB} = \Xi V \Xi + (1-\Xi^2)N_T,
\end{equation}
where $\Xi = \mathrm{diag} [(1,1,\sqrt{\eta},\sqrt{\eta})]$ and $V$ is taken from (\ref{varmatrix}). However, the global state $\rho_{ABE}$ is pure - all the noise and loss present in the state, which is shared by Alice and Bob, represents information gained by Eve. After Bob performs his measurement, obtaining value $b$ with probability $P(b)$ and leaving the Alice's and Eve's modes in the state $\rho_{AE}(b)$, the information about Alice's initial state gained by Eve can be upper-bounded by Holevo quantity \cite{holevo}:
\begin{eqnarray}\label{}
    H(E;A) &=& S(\rho_E) - \int P(b) S[\rho_E(b)] db \nonumber \\ &=& S(\rho_{AB}) - \int P(b) S[\rho_A(b)] db.
\end{eqnarray}
Here $S(.)$ stands for von Neumann entropy, $\rho_E = \Tr_{AB}[\rho_{ABE}]$, $\rho_E(b) = \Tr_A[\rho_{AE}(b)]$, $\rho_A(b) = \Tr_E[\rho_{AE}(b)]$, and we have taken advantage of purity of $\rho_{ABE}$ and $\rho_{AE}(b)$. For the case of reverse reconciliation, where Eve's knowledge about Bob's measurement result is given by $H(E;B)$, similar relation can be obtained by interchanging variables corresponding to Alice and Bob,
\begin{eqnarray}\label{}
    H(E;B) &=& S(\rho_E) - \int P(a) S[\rho_E(a)] da \nonumber \\ &=& S(\rho_{AB}) - \int P(a) S[\rho_B(a)] da,
\end{eqnarray}
where $\rho_E(a) = \Tr_B[\rho_{BE}(a)]$, $\rho_B(a) = \Tr_E[\rho_{BE}(a)]$, and $\rho_{BE}(a)$ represents the state of the system after Alice's measurement yielded value $a$ with probability $P(a)$.

The purity of the global state is an important assumption, which does not mesh well with the noise-powered amplification routine. If the amplification noise was simply added, it would lead to overestimation of Eve's information. Therefore, we are going to represent the noise addition in a different manner. Taking hints from the virtual entanglement approach, we introduce an ancillary state $|\nu\rangle = \sum_{k = 0}^{8}c_k |k\rangle$ to which corresponds a set of coherent amplitudes ${\beta_k}$ such that $\beta_0 = 0$ and $\beta_k = \sqrt{\Delta'}e^{i2k\pi/8}$. The coefficients $c_k$ can be arbitrary, but we have chosen $c_0 = 1$ and $c_k = e^{-\Delta'/2}$ for $k = 1,\ldots 8$ (prior to normalization) to emulate Gaussian distribution of thermal noise. This state is under Bob's control and the process of noise addition can be now described by
\begin{equation}\label{}
    \rho \rightarrow \sum_{k=0}^8 D(\beta_k)\otimes |k\rangle\langle k| (\rho\otimes|\nu\rangle\langle\nu |)\sum_{k=0}^8 D(-\beta_k)\otimes |k\rangle\langle k|.
\end{equation}
For the purpose of obtaining the mutual information between Alice and Bob, the ancillary mode $\nu$ is traced over. For the purpose of obtaining Eve's information, Bob is considered measuring both modes B and $\nu$. The noise modeled in this way is non-Gaussian, but because we consider the noise addition procedure to be under our control, there are no limits to the actual form the noise might take. In fact, it was even demonstrated that non-Gaussian noise is beneficial for some kinds of applications \cite{NPcloning}.

The analysis has been again performed for a single quantum state (two mode squeezed state with $R = 0.3$) used in four different communication protocols represented by combinations of measurements Alice and Bob can perform. Table~\ref{holevo_comparison} shows the differences of the relative quantities before the operation and after. Positive and negative values correspond to the values being increased and decreased, respectively, by the amplification operation.

The benefits of amplification are best visible in scenarios in which Alice and Bob's measurements match - they both perform either homodyne (squeezed state communication), or heterodyne (coherent states communication) detections. The number of scenarios is doubled, though, by considering both direct and reverse reconciliation. The further analysis shall be therefore concerned only with those scenarios. Four different Gaussian channels have been analyzed and the results can be seen in Table~\ref{holevo_comparison}.

The first channel is again the ideal one, with no losses or added thermal noise. In such the channel Eve has no information and this remains unchanged under amplification, which still does increase the mutual information between Alice and Bob, as was discussed in section \ref{Sec3}. The second type of channel is represented by added noise of $N_T = 0.1$ but no loss. In this case, both amplifiers across all the four communication protocols improve Alice's and Bob's situation over Eve's. It is interesting now, to look at the effect of added noise. For two photon subtraction, adding noise both improves $I(A,B)$ and (most of the time) reduces $H(E;.)$. As a consequence, noise is necessary if the amplification is to have positive impact for direct reconciliation protocols. For addition and subtraction of a photon, the noise does not improve $I(A,B)$ at all, as could be expected from the analysis of the pure case. However, the noise hampers Eve even more - so much, in fact, that for reverse reconciliation protocols it is significantly more beneficial than the pure scenario.

The third channel is pure lossy channel with $\eta = 0.9$, with no thermal noise. As such, it is very similar to the ideal noiseless and lossless channel in it's capacity to accommodate amplifiers - both of them lead to improvements for all communication protocols, with artificial addition of noise being useful for two photon subtraction and detrimental for photon addition and subtraction.

The last considered channel exhibits both noise $N_T = 0.1$ and loss $\eta = 0.9$. As such it is the scenario with strongest ties to reality. What distinguishes it from the other scenarios most, is the comparatively huge improvement in information Eve gains when Alice and Bob attempt any amplification. It is so significant, in fact, that for communication via coherent states (heterodyne-heterodyne) the NP amplifier fails to achieve any improvement. It can, however, be helpful in the case squeezed state communication (homodyne-homodyne) with reverse reconciliation. In this case, the artificially added noise is not necessary, but helpful. On the other hand, the HF amplifier is helpful only for the direct reconciliation protocol and for those the addition of noise is only detrimental. Interestingly enough, when enough noise is added, addition and subtraction of photon can also lead to improvement for reverse reconciliation in the squeezed state communication.

Let us pause now and attempt to summarize the effect of artificial added noise. If addition of noise precedes two photon subtraction, it generally increases mutual information between Alice and Bob. However, at the same time it also increases (for reverse reconciliation) or decreases (for direct reconciliation) the information of Eve. On the other hand, for addition and subtraction of photon, addition of noise always hurts Alice and Bob, but it also always reduces the influence of Eve. As a consequence, for both types of amplifier there are scenarios, when addition of noise is beneficial, and even scenarios, when is it essential for improvement to appear.

\section{Summary}\label{Sec5}
We have analyzed the effect noiseless amplifiers can have on quantum communication. Specifically we were concerned with two types of amplifier: HF amplifier consisting of photon addition and subtraction, and NP amplifier realized by addition of noise and photon subtractions. Special attention was devoted to the effect of artificial noise addition - it's influence was also studied for the HF amplifier, even though the original design does not call for it.

First we have studied the amplifiers purely from the perspective of mutual information between the communicating parties. It was revealed, that both amplifiers can improve the mutual information. The effect of artificial noise differs, though. For the HF amplifier the noise is always detrimental. For the NP amplifier, however, it improves the situation almost to the level of the HF amplifier - addition and subtraction of photon is significantly better than two photon subtraction (hinting towards photon addition being more `powerful' operation than photon subtraction), but roughly at the same level as two photon subtraction after noise addition.

In the second step we have analyzed the amplifiers within the context of a full communication protocol, with eavesdropper Eve included. To have comparison between several physical scenarios, we have studied four different quantum channels: ideal channel, purely noisy channel, purely lossy channel, and realistic noisy and lossy channel. For channels without any added noise, both amplifiers work reliably and without surprises - the performance is mostly given by changes in mutual information, which is increased by the amplification for all communication scenarios. For the purely noisy channel, both amplifiers lead to improvement, but surprisingly, even the performance of the HF amplifier can be improved by adding artificial noise. If the channel is lossy and noisy at the same time, the regimes in which the amplifier is beneficial are most difficult to find, but it can be done. And again, it may be advantageous, for both types of amplifiers, to add a measure of artificial nose.

We have shown that realistic experimentally feasible amplifiers can be used to improve existing quantum information protocols, and that addition of artificial noise can play an important role in this task.

\medskip
\noindent {\bf Acknowledgments} We acknowledge support from grant P205/12/0577 of Grant Agency of Czech Republic.

\end{document}